\documentclass[aip,reprint, apl]{revtex4-2}
\usepackage{graphicx}
\usepackage{dcolumn}
\usepackage{bm}
\usepackage{etoolbox}
\usepackage[utf8]{inputenc}
\usepackage[T1]{fontenc}
\usepackage{mathptmx}
\usepackage{textcomp, gensymb}
\usepackage{xcolor}
\usepackage{amsmath} 
\newcommand\note[1]{\textcolor{black}{#1}}

\begin{document}


\title{Knife-Edge Diffraction of Scalar and Vector Vortex Light} 



\author{Richard Aguiar Maduro}
\affiliation{School of Physics \& Astronomy, University of Glasgow, Glasgow G12 8QQ, UK}

\author{Amanda Kronhardt Fritsch}
\affiliation{School of Physics \& Astronomy, University of Glasgow, Glasgow G12 8QQ, UK}
\affiliation{Instituto de Física, Universidade Federal do Rio Grande do Sul, Av. Bento Gonçalves 9500 Porto Alegre, RS, Brazil}

\author{Sonja Franke-Arnold}
\email[]{sonja.franke-arnold@glasgow.ac.uk}
\affiliation{School of Physics \& Astronomy, University of Glasgow, Glasgow G12 8QQ, UK}


\date{\today}

\begin{abstract}
Various methods have been introduced to measure the orbital angular momentum (OAM) of light, from fork holograms to Dove prism interferometers, from tilted lenses to triangular apertures - each with their own benefits and limitations. Here we demonstrate how simple knife-edge diffraction can be used to identify the OAM of a optical phase vortex from the formation of fork dislocations within the Fresnel diffraction pattern. For vector vortex beams without net OAM, the conventional Fresnel fringes are recovered, whereas the polarization in the geometric shadow is changed in its ellipticity. The observed diffraction patterns agree with simulations and their features can be explained by considering diffraction as an interference phenomenon. Knife-edge diffraction provides not only an instructive illustration of various properties of phase and polarization vortices, but can also serve as an ideal method for the quick determination of optical OAM, with potential applications beyond optics, where alternative detection measurement methods may be harder to realize. 
\end{abstract}

\pacs{}

\maketitle 


\section{Introduction}

Light provides a canvas that may be structured in its intensity, phase, and polarization among other degrees of freedom. Since the beginning of this millennium, the arrival of spatial light modulators, digital mirror displays and other specialized devices in optical laboratories has allowed us to generate and analyze a wide range of optical structures. Optical vortices, be it phase vortices carrying optical angular momentum (OAM) within homogeneously polarized light, or vector vortices featuring polarization singularities, capture the attention of researchers more than other optical structures.

Phase vortices are characterized by a helical phase structure that twists around the optical axis, resulting in a phase singularity (or dislocation) in the center of the beam \cite{Allen1992,Beijersbergen1994, ALLEN1999291, SOSKIN2001219, bekshaev2008paraxial, DENNIS2009293}. They have been the subject of numerous investigations, summarized in various reviews  \cite{Yao:11,RubinszteinDunlop2016,Forbes2023, FrankeArnold2022, Cisowski2022}.

Vector light, characterized by inhomogeneous polarization profiles, is a more recent addition to the structured light family. Vector light may feature vector vortices with topological singularities in the polarization field \cite{Zhan:09}. Typically, these arise when superimposing different phase vortices in orthogonal polarization components. For details on their characterization as well as their significance for light matter interaction and optical information encoding we refer the reader to recent reviews \cite{Wang2020,Forbes2021,Shen2022,Lv2024}.

While the interference of phase structures results in intensity modulations, the interference of polarization structures results in polarization modulations across the beam profile. The presence of optical vortices is therefore visible in any experimental situation that involves interference in its widest sense, including e.g.~diffraction or scattering. This needs to be considered when interpreting measurements, and in turn may be utilized for the detection of vortex light.

Direct interference with its mirror image or a reference beam was perhaps the first method identified to measure the OAM of a light beam \cite{Harris1994, Basistiy:95} -- and the characteristic spiraling or fork-like intensity patterns are still widely used \cite{Lin:18}. Even single plano-concave lenses can \note{serve} 
as rudimentary interferometric devices by superposing the reflections from the front and back surfaces of the lens \cite{Cui:19}. Mode-conversion may be understood as a less obvious form of (intra-beam) interference, and \note{its use for the analysis of vortex light includes} 
astigmatic systems (such as tilted or cylindrical lenses) \cite{Courtial:99, Vaity:13}, as well as more sophisticated devices that work on the single photon level \cite{Berghout2010, Leach2006}. 
Diffraction offers an alternative route, including multiple-pinhole diffraction \cite{Berkhout:08}, single or double-slit diffraction \cite{GHAI2009123, Sztul:06, Ferreira2011,Narag:19}, and diffraction through triangular or circular apertures \cite{Hickmann2010, Mourka:11, Ambuj:14}, as well as different types of gratings \cite{Dai:15, Liu:18, Zhang:18, Hebri:18}, single-point detection \cite{Li:18}, spatial light modulators \cite{Forbes:16}, and specially designed nanoholes \cite{Jin2016}. \note{In this paper we will elaborate on knife-edge diffraction as an efficient and experimentally simple topological method to identify the charge and sign of scalar vortices, and furthermore to illustrate features of vector vortices.}

Knife-edge diffraction can be understood as the interference of two superimposing waves: the geometrical wave from the input beam and the boundary diffraction wave from the diffracting object. It has been beautifully demonstrated that this effect can be used to analyze the phase structure of a diffractive object \cite{Kumar2007}. In reverse, diffraction off a known object ({\it i.e.}~the knife-edge) \note{can reveal critical information about the light beam's phase structure \cite{Bekshaev2018}}. Two decades ago, diffraction of a phase vortex by a knife-edge has been used to probe the handedness of its topological charge \cite{Arlt2003}. Like other single-beam interferometers,
knife-edge diffraction is self-referencing, and hence provides
some stability to aberrations and external vibrations. \note{The sign and charge of an optical phase vortex is revealed by the orientation of the fork dislocation present in the diffraction pattern, making knife-edge diffraction an experimentally simple topological method for the analysis of optical vortices.}

In this paper we present a detailed experimental study of knife-edge diffraction of phase and polarization vortices. We find that the typical Fresnel fringes are modified for light that carries net OAM, generating fork dislocations, which can be used to identify both the charge and sign of the topological charge. 
For polarization vortex beams without net OAM, the original Fresnel fringes are recovered, but polarization tomography reveals a modification of the input polarization pattern, which can be explained by considering the different Poynting vectors of the orthogonal polarization components. Our experimental results compare well with simulations based on the angular spectrum method.

\section{Theoretical description}

\subsection{Phase and polarization vortex beams}
Point singularities in the electric field may arise at positions of undefined phase and/or polarization, and are characterized by their topological charge (winding number). Phase singularities of topological charge $m$ are associated with an azimuthal phase structure $\exp(i m \phi)$, responsible for an OAM of $\hbar m$ per photon. For polarization singularities, it is the winding of the Stokes phase around the point of indeterminate polarization that determines the topological charge\cite{Zhan:09}.

While vector light offers a rich and versatile repertoire of optical structures, in this paper we restrict ourselves to the simplest examples of phase and polarization vortices, namely to paraxial vector fields of the form
\begin{equation}
\label{PV}
\Vec{u}(r,\phi) = \frac{1}{\sqrt{2}} \left( {\rm LG}_0^{m} e^{i\alpha/2}\hat{l} +  {\rm LG}_0^{-m}e^{-i\alpha/2} \hat{r} \right).  
\end{equation}
Here $r$ and $\phi$ denote the radial and azimuthal position within the beam profile, and we have denoted the beam in terms of the (right and left) circular polarization basis, defined as  $\hat{r} = (\hat{h} -i\hat{v})/\sqrt{2}$ and $\hat{l} = (\hat{h} + i\hat{v})/\sqrt{2}$ where $\hat{h}$ and $\hat{v}$ represent horizontal and vertical polarization directions, respectively, and $\alpha$ denotes the phase difference between the orthogonal polarization components. 
The spatial modes associated with the right and left polarization components are Laguerre-Gauss functions 
\begin{eqnarray} \label{LG}
     {\rm LG}_{p}^m (r,\phi,z) & = & \frac{C_{pm}}{w_{0}}\left(\frac{\sqrt{2}r}{w(z)}\right)^{|m|} L_{p}^{|m|}\left(\frac{2r^{2}}{w(z)^{2}}\right) \exp\!\left(\!-\frac{r^{2}}{w(z)^{2}}\!\right)\nonumber \\ 
     & &  \exp(i m\phi) \exp(i \Phi_\mathrm{G}) \exp\left[i \left(k- \frac{r^2 }{w(z)^2 z_\mathrm{R}}\right)z\right] 
\end{eqnarray}
propagating along the z direction, with $w(z)$ denoting the beam waist and $\Phi_\mathrm{G} =(2p+|m|+1)\arctan(z/z_\mathrm{R})$ the Gouy phase. Here $z_\mathrm{R} = \pi w_0^2/\lambda$ is the Rayleigh range, $L^{|m|}_{p}$ is the associate Laguerre polynomial, and $C_{pm}$ a normalization constant. For greater clarity, we concentrate on the azimuthal structure of the light and assume the radial mode number $p$ to be zero.

While Eq.~(\ref{PV}) describes a polarization vortex of charge $2m$, its individual left and right handed represent phase vortices. They are given by ${\rm LG}_{p}^{\pm m} \propto \exp(\pm i m \phi)$ and hence associated with an OAM of $\pm \hbar m$ per photon. These are associated with a tilted Poynting vector with a skew angle $\propto \nabla u(r,\phi) =\pm m/(k r)$, indicating that the direction of energy flow between the right and left hand component is reversed. 
The complete vector beam instead has no net OAM, and its Poynting vector is directed along the beam propagation direction. 

\begin{figure}    \includegraphics[width=\columnwidth]{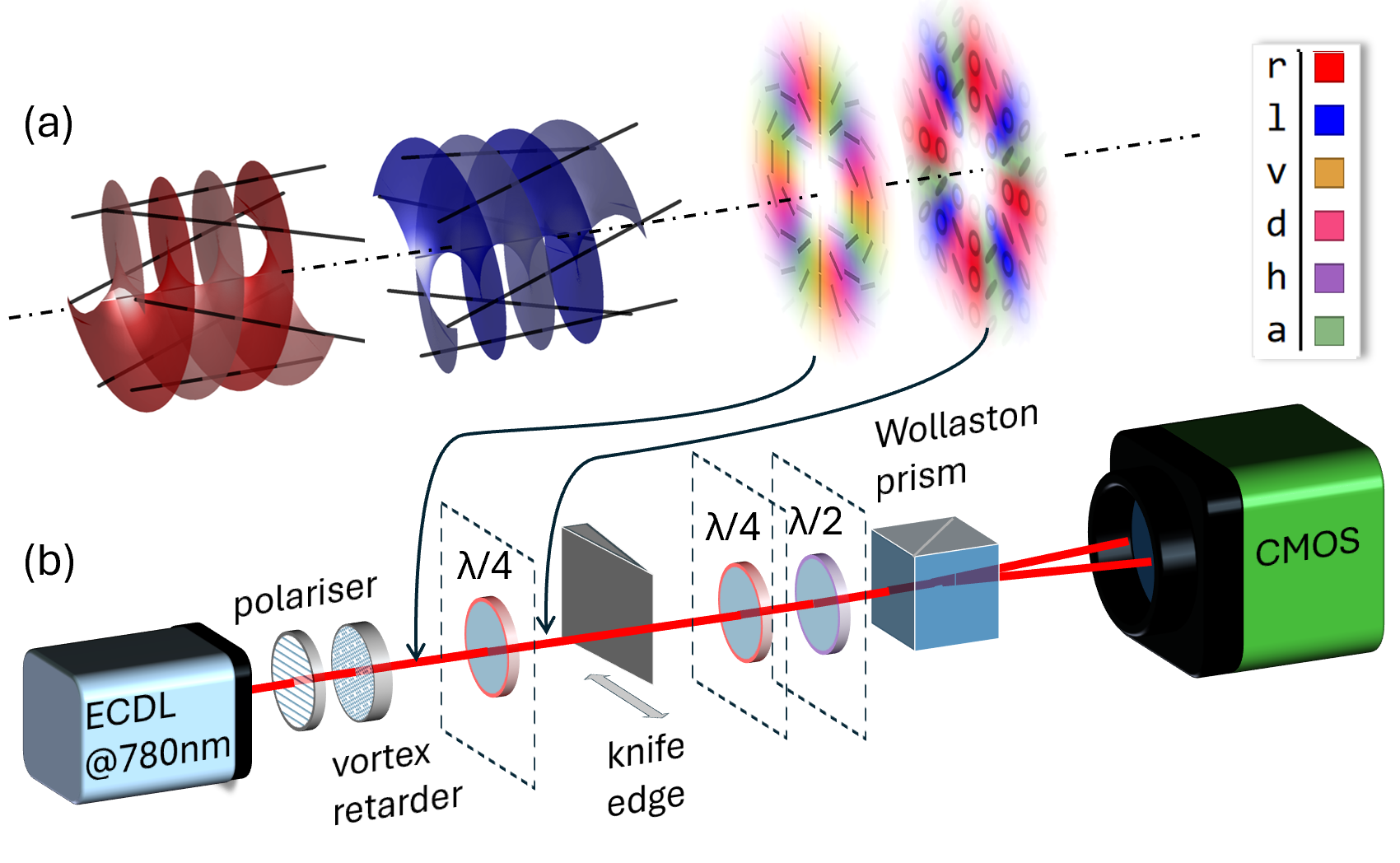}
    \caption{Knife-edge diffraction of vortex beams. (a) Illustration of a vector vortex beam, composed as a superposition of a right circular polarization component (red) with $m=-2$ and a left circular polarization component (blue) with $m=+2$. The corresponding polarization profile is shown to the right, followed by the profile after passing through a QWP. Inset: colorscheme used to display the polarization states.  
    (b) Simplified experimental setup, including beam preparation, knife edge and additional polarization optics for Stokes polarimetry. The functions of the various optical elements are explained in the text.}
    \label{fig:setup}
\end{figure}

An example of a vector vortex beam of the form of Eq.~(\ref{PV}) is illustrated in Fig.\ref{fig:setup}(a) for $m=\pm2$, showing the twisted wavefronts and Poynting vectors of the right and left polarization component as well as the resulting polarization vortex. We will discuss the generation of such light in section \ref{s:meth}.  

\subsection{Simulation}
The Angular Spectrum Method (ASM) is used to calculate field diffraction numerically  \cite{goodmanlivro, CJB, ELalor, DOPC}, providing predictions and comparisons for our measurements. 
It allows us to calculate the optical field of a propagating beam at any plane, modeling both free propagation and diffraction caused by obstacles.  

We consider forward propagation along $z$ in the paraxial approximation. The ASM can be summarized in terms of two-dimensional Fourier transforms:
\begin{equation}
U(x,y;z)=\mathcal{F}^{-1}\{\mathcal{F}\{U(x,y;0)\}\times H(f_x,f_y;z)\},
\end{equation}
where $H(f_x,f_y;z)=e^{2\pi i\left(f_xx+f_yy+\frac{z_0}{\lambda}\left(\sqrt{1-(\lambda f_x)^2-(\lambda f_y)^2}\right)\right)}$ is the spatial frequency transfer function and $\mathcal{F}^{-1}$ is the inverse Fourier transform \cite{poon}.
We implement our simulation using MATLAB software.
As diffraction is, to first order, a linear effect, the propagation of each polarization component can be modeled independently. The propagated total vector field is obtained by the superposition of the orthogonal polarization components, allowing us to reconstruct the intensity, phase and polarization profile of vector vortices diffracted by a knife edge.

We model the propagation and diffraction of Eq.~(\ref{PV}) based on ideal LG modes. 
The real-space grid is defined based on the experimental parameters of the imaging system, and the corresponding reciprocal space grid is made sufficiently large to allow propagation in the Fourier domain.
The knife-edge is modeled as a binary mask, where regions of the beam beyond a specified position (defined in terms of the beam waist $w_0$) are blocked. 
The beam is also confined by a circular aperture to simulate the finite size of optical elements.

The polarization state of the optical field is analyzed using Jones calculus formalism. The electric field components \((E_x, E_y)\) of the beam are determined for superpositions of LG beams with orthogonal polarizations (e.g., right- and left-circular polarization). From these components, the Stokes parameters
\begin{equation} \label{Stokes} \mathbf{S}=
    \begin{pmatrix} S_0\\S_1\\S_2\\S_3\end{pmatrix}=\begin{pmatrix} I_\textrm{h}+I_\textrm{v} \\I_\textrm{h}-I_\textrm{v}\\I_\textrm{d}-I_\textrm{a}\\I_\textrm{r}+I_\textrm{l}\end{pmatrix} =\begin{pmatrix} |E_x|^2+|E_y|^2\\|E_x|^2-|E_y|^2\\E_x^*E_y+E_x^*E_y\\-i(E_x^*E_y+E_xE_y^*)\end{pmatrix},        
 \end{equation}
 are calculated to fully characterize the polarization state. The Stokes vectors vary as a function of the transverse position within the beam profile, which for simplicity we have omitted in Eq.~(\ref{Stokes}). For completeness we include the description in terms of the intensities in the horizontal, vertical, diagonal, antidiagonal, right and left circular polarization component, $I_\textrm{h}, I_\textrm{v}, I_\textrm{d}, I_\textrm{a}, I_\textrm{r}$ and $I_\textrm{l}$.  

The simulation results for intensity, phase and polarization patterns \note{offer a numerical validation of the experimental measurements, serving as a qualitative confirmation of the interpretation of this phenomenon through the various visualizations and analytical tools given in the following section \ref{s:meth}}. While the intensity patterns reveal the spatial distribution of optical power, phase maps highlight the structure of vortices, and polarization maps provide insight into the interplay between spatial truncation and polarization states. These visualizations are key to understanding the impact of truncation on the beam's propagation and diffraction properties.
\section{Experimental setup and results \label{s:meth}}

We demonstrate the effects of knife-edge diffraction of both phase and polarization vortex beams using the setup shown in Fig.~\ref{fig:setup}(b). 

\subsection{Experimental setup}
A $780$nm laser beam, generated by a home-made external cavity diode laser (ECDL) and powered by a Moglabs diode laser controller DLC202, is collimated and vertically polarized. The laser beam is spatially filtered by passing it through a single-mode fiber followed by a telescope system to generate input light close to the $\mathrm{LG}_0^0$ mode with a waist size of $w_0=4.1$mm. 
The laser beam passes through a vortex retarder with a given topological charge $m$, generating a superposition of circularly polarized polarization components with opposite OAM as given in Eq.~(\ref{PV}),\footnote{Strictly speaking, the vortex retarder only imposes the required azimuthal phase profile onto the input beam, but after propagation (not shown) this resembles the desired $\mathrm{LG}_0^m$ profile.} creating a vector vortex beam with a spatially varying polarization pattern, as indicated in Fig.~\ref{fig:setup}(a) for the case of $m=\pm 2$ . Higher order vortex beams can be generated by combining two vortex retarders with values $m_1$ and $m_2$ with a half waveplate (HWP) placed between them, creating a vortex of charge $m_1+m_2$ \cite{Delaney2017}. 

In order to access the constituting phase vortices more easily, we transform this beam from the circular to the Cartesian polarization basis: Propagation through a further quarter wave plate (QWP) at $45\degree$  transforms this beam into
\begin{equation}
\mathbf{u}(r,\phi) = \frac{1}{\sqrt{2}} \left(e^{i\alpha/2}{\rm LG}_0^m \hat{h}+e^{-i\alpha/2}{\rm LG}_0^{-m} \hat{v}\right), 
\end{equation} 
whose polarization profile is indicated at the right of Fig.~\ref{fig:setup}(a). This form of vector vortex beam allows us to extract the phase vortices by projection onto the horizontal and vertical polarization basis using a Wollaston prism. A vertically positioned knife-edge intersects this hybrid vector beam, and its intensity profile can be monitored on a jAI CMOS camera, positioned in the far field of the knife edge. 
We note that alternatively, we could have inserted the Wollaston prism {\it before}, but as diffraction is a linear process, the action of the Wollaston prism and the knife edge are interchangeable. 
Removing the Wollaston prism and performing spatially resolved Stokes tomography instead allows us to investigate the diffraction of the polarization vortex.

We remove unwanted interference fringes from the light's interaction with the surface of the camera by applying a simple Fourier filter: a fast Fourier transformation (FFT) is performed on the raw image, and peaks corresponding to interference fringes are removed with a suitable mask. 
Finally, an inverse Fourier transformation function returns the filtered experimental intensity profile. 

\subsection{Knife-edge diffraction of phase vortex beams}

\begin{figure}[!t]
\centering
\includegraphics[width=\columnwidth]{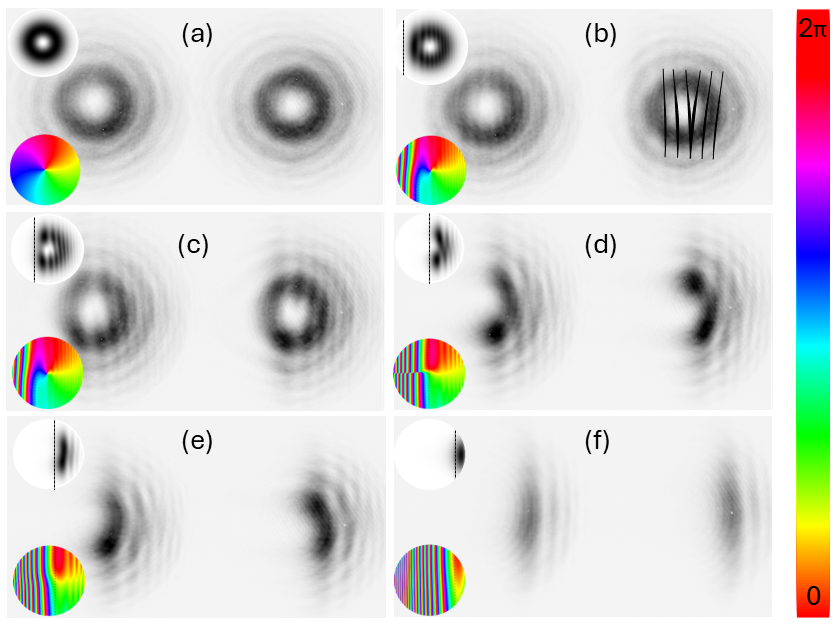}
    \caption{Observed diffraction pattern for light with $m=+1$ (left image) and $m=-1$ (right image) as a function of the lateral position of the knife edge at positions (a)$-2w_0$, (b) $-w_0$, (c) $-w_0/2$, (d) 0, (e) $w_0/4$, (f) $w_0$, with respect to the beam's center. The circular insets show the corresponding simulations for $m=+1$, with the intensity profile displayed in gray levels, the phase in hue colors, and the dotted lines showing the edge position (those for $m=-1$ are horizontal mirror images).\note{To guide the eye, we have indicated the ``fork'' with gray lines in (b).}
  }
\label{fig:knife_prog}
\end{figure}

We first discuss our observations of the knife-edge interference fringes when moving the lateral position of the knife-edge through phase vortices of charge $m=1$ and $m=-1$. The diffraction patterns recorded on the CMOS sensor, are shown in Fig.~\ref{fig:knife_prog} for a selection of lateral positions.

Before the knife-edge intersects the beam significantly, the $m=\pm 1$ phase vortices are intact as shown in Fig.~\ref{fig:knife_prog}(a), but as the knife intersects the light (b-c), the characteristic fork pattern emerges. The diffraction patterns for $\pm m$ are horizontal mirror images of each other, featuring a fork dislocation with \note{$|m|+1=2$ prongs}. The sign of the vortex can clearly be identified from the fork orientation, confirming a previous study \cite{Arlt2003}. For $m=+1$ ($m=-1$) more light invades the geometric shadow in the lower (upper) half plane, in agreement with the clockwise (anticlockwise) \note{energy circulation\note{\cite{Bekshaev2014}} as determined by the} tilt of the Poynting vectors with skew angles $\pm 1/(kr)$. 
Once the knife edge passes the vortex position (e,f), the fork disappears again, however the tilt of the diffraction pattern remains. 

The insets in Fig.~\ref{fig:knife_prog} show simulations of the corresponding intensity and phase maps. 
The intensity maps are peak-normalized, and the phase distributions are displayed as HUE color maps for enhanced visualization. The phase in the region of the geometric shadow shows a rapid phase variation, associated with the fact that the interference pattern may be understood as the superposition of the direct beam and the one originating from the diffracting object.
Before the knife edge reaches the vortex position, Fig.~\ref{fig:knife_prog}(a-c), the point singularity of the vortex remains intact. As the knife edge reaches the vortex, Fig.~\ref{fig:knife_prog}(d), this is converted to a horizontal line singularity in the geometric shadow region, which disappears once the knife-edge obstructs the vortex. 

\begin{figure}[!t]
\includegraphics[width=.8\columnwidth]{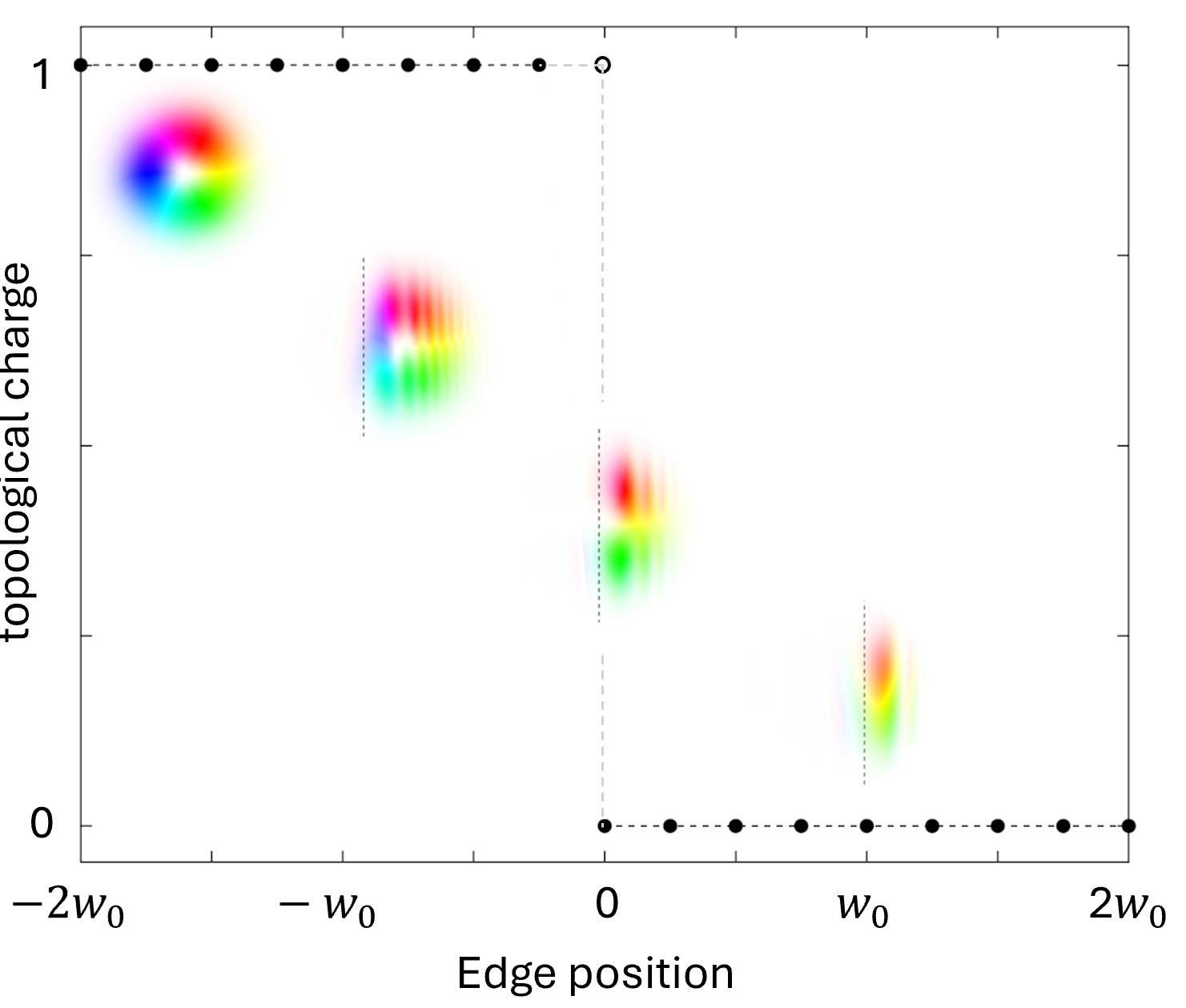}
\caption{Topological charge of the diffracted beam as a function of the edge position relative to the singularity, evaluated from phase simulations (full circles). The plots for positions $-2 w_0$, $-0.75 w_0$, $0$, and $0.75 w_0$ illustrate the emergence of the fork diffraction pattern and its subsequent disappearance as the edge moves beyond the beam's center.}
    \label{fig:phasedifference}
\end{figure}

We analyze the presence and disappearance of the point vortex in more detail in Fig.~\ref{fig:phasedifference}. 
This figure displays the accumulated phase along a closed path around the beam center, {\it i.e.} the singularity position of the input beam, divided by $2\pi$. This value measures the vortex charge, corresponding to the \note{OAM of the input beam \footnote{We note that, as the propagation direction changes during diffraction, the OAM or the input light does not coincide with that of the diffracted beam - for small diffraction angles considered here it is however a good approximation.}}. 
As expected, the phase initially exhibits a \( 2\pi \) variation around the vortex (for \( m=1 \)). As the edge covers the vortex, the accumulated phase goes to zero, corresponding to the disappearance of the fork-like fringes in Fig.~\ref{fig:knife_prog}. Analogous behavior occurs also for negative and higher-order vortices (not shown).

\begin{figure}[!b]
\centering
\includegraphics[width=0.95\columnwidth]{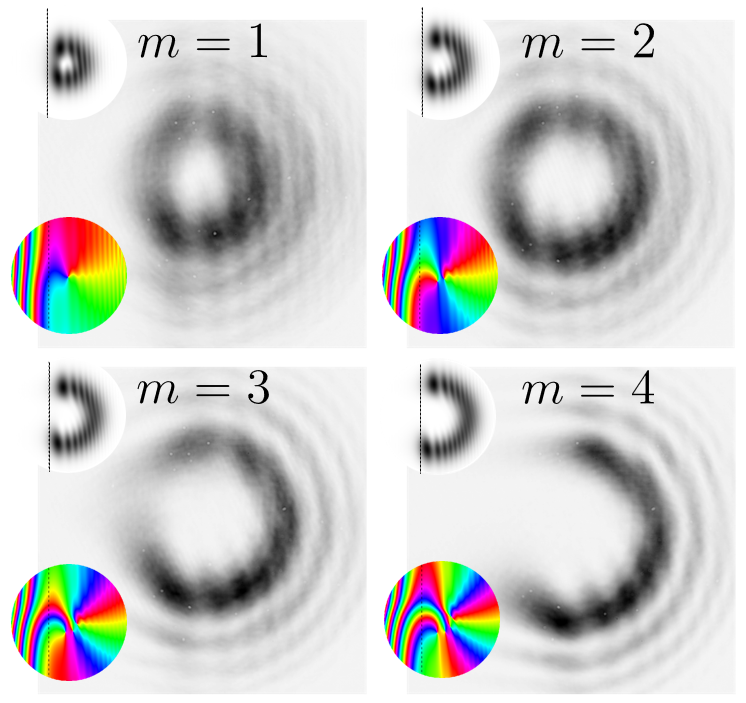}
    \caption{Knife-edge diffraction patterns for phase vortex beams with $m=1$, 2, 3 and 4 with the knife edge at $x=-w_0/2$ (as in Fig.\ref{fig:knife_prog}(c)), revealing an $m$-pronged fork. The corresponding patterns for beams with negative OAM (not shown) are horizontal mirror images. The corresponding simulations of intensity and phase follow the format of Fig.~\ref{fig:knife_prog}. }
    \label{fig:top_charge}
\end{figure}

We have also measured and simulated knife-edge diffraction for higher order phase vortices, as illustrated in Fig.~\ref{fig:top_charge}, choosing a lateral knife-edge position of $-w_0/2$ which gives the clearest fork patterns. 
\note{For an input beam with topological charge $m$,} we observe \note{$|m|+1$}-pronged forks.
\note{Together with our results from Fig.~\ref{fig:knife_prog},}  
this demonstrates that knife-edge diffraction is a simple and effective method to determine the value and sign of the topological charge of a vortex beam \note{from the shape and orientation of the fork dislocation.} 
Our measurements also indicate that the extent to which the diffracted beam penetrates the geometric shadow region is influenced by \( |m| \). Again, we may understand this intuitively by considering the tilt of the Poynting vector in the presence of a vortex, which is proportional to \( |m| \). This phenomenon can be interpreted as a redistribution of energy due to the vortex topology, which influences the diffraction pattern's behavior \cite{Berry2002, Soskin1997, Gbur2005, Bekshaev2014}.

\subsection{Knife-edge diffraction of polarization vortex beams}

\begin{figure}[!b]
     \centering     \includegraphics[width=\columnwidth]{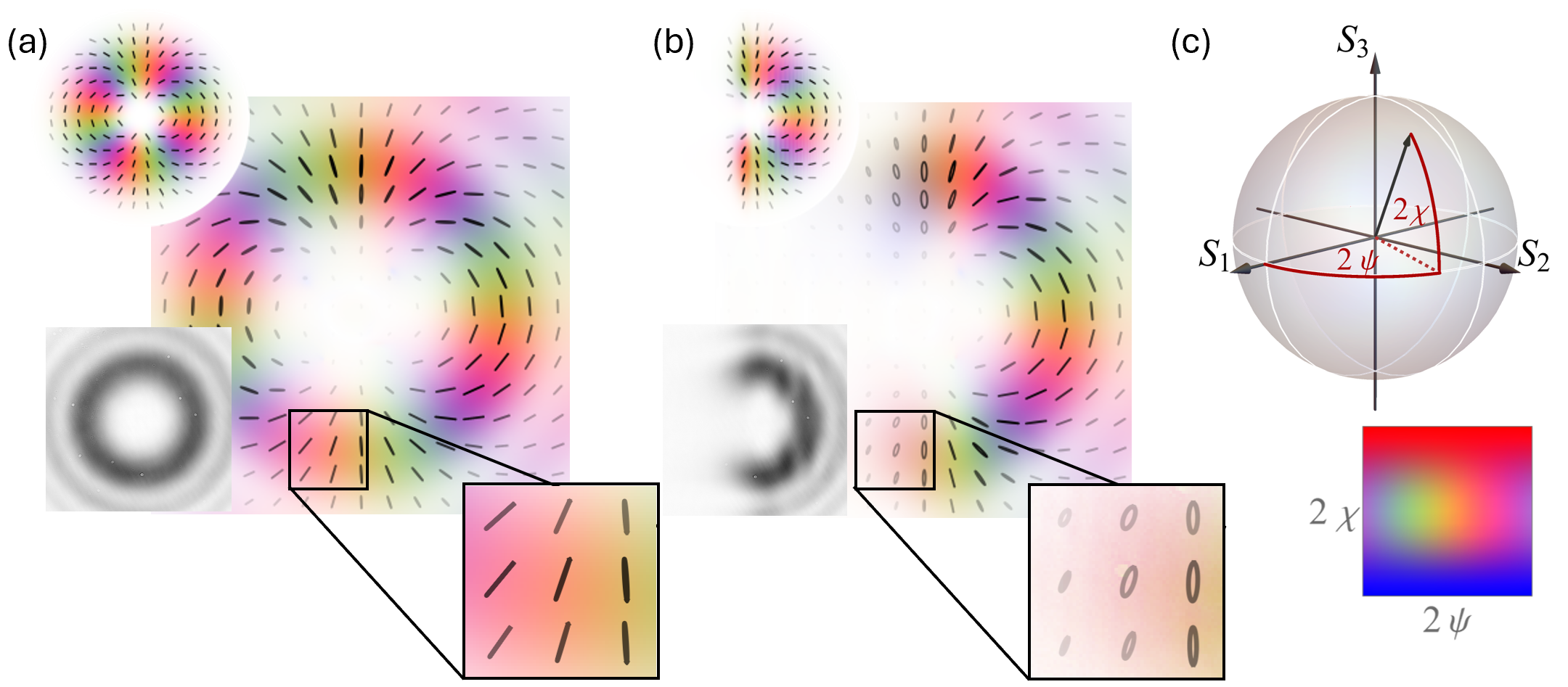}
     \caption{Knife-edge diffraction of a polarization vortex beam. The polarization profile of the input beam 
      $\left(\mathrm{LG}_0^{-2} \hat{r}-\mathrm{LG}_0^{2}\hat{l}\right)/\sqrt{2}$ is shown in (a) and that of the beam diffracted from a knife-edge at $-w_0/3$ of the beam profile in (b), showing predominantly left/right handed polarization at the top/bottom shadow regions. The measured intensity profiles are added as gray-level insets, and simulations as circular insets. (c) details the color scheme used to represent the relation between ellipticity $2\chi$ and the orientation $2\psi$ on the Poincaré sphere.}
     \label{fig:Polarization}
\end{figure}

In order to investigate the effect of knife-edge diffraction on polarization vortices, we remove the QWP and restore the vortex beam of Eq.~(\ref{PV}), based on spatially varying linear polarizations.  The interference pattern visible on the CMOS camera shows the typical knife-edge fringes expected for light without OAM (shown as square insets in Fig.~\ref{fig:Polarization}). In order to capture the change in polarization across the diffracted beam, we perform full Stokes tomography by measuring its intensities in the various polarization basis, as indicated in Eq.~(\ref{Stokes}). Experimentally this is achieved by placing additional half and quarter-wave plates in front of the Wollaston prism. The resulting locally varying Stokes vector $\mathbf{S}(\mathbf{r}_\bot)$ is displayed by a grid of corresponding polarization ellipses with spatially varying orientation and degree of ellipticity. In addition, polarization distributions are visualized using color-coded maps that represent the intensity, degree of ellipticity ($-\pi\leq 2\chi \leq \pi$), and orientation of the electric field vector ($0 \leq 2\psi \leq 2 \pi$). This allows for a detailed analysis of how spatial truncation and propagation affect the beam's polarization structure.



Fig.~\ref{fig:Polarization} shows the measured and simulated polarization patterns of the vector vortex beam in Eq.~\ref{PV} for $m=2$, with the unperturbed beam shown in Fig.~\ref{fig:Polarization}(a) and the beam after a knife-edge at position $-w_0/3$ in Fig.~\ref{fig:Polarization}(b). Both data and simulation show that the purely linear polarizations of the input light acquire a slight degree of ellipticity around the location of the knife edge. In the upper half plane, the ellipticity is right-handed (indicated by a reddish tint), and in the lower half plane left-handed (indicated by a blueish tint). This modulation of the polarization pattern can again be explained from the decomposition of the vector vortex beam into its right and left-handed components, associated with opposite OAM values and hence oppositely tilted Poynting vectors, in agreement with our previous observations.    



\section{Conclusion}
Knife-edge diffraction does not receive the same attention as other diffraction phenomena like the double slit or gratings, despite its honorable mention in undergraduate optics courses and its \note{impact on} wireless communication. Here we have presented a detailed analysis of knife-edge diffraction for scalar and vector vortices. For phase vortices, we observe fork dislocations within the familiar Fresnel diffraction fringes that depend on both the sign and magnitude of the vortex charge. Our observations agree with simulations based on the ASM. For vector vortices, we find a modulation of the ellipticity close to the shadow region. The effects in both scalar and vector vortices illustrate the energy flow given by the tilted Poynting vector of the phase vortex beams. 

While our experiment was optimized to demonstrate the fork dislocations in the knife-edge diffraction pattern as cleanly as possible, a qualitative observation of the diffraction pattern can be achieved very quickly by inserting a blade partly into a laser beam, offering a simple and effective method to determine the charge and sign of a vortex. \note{Counting dislocations of a fork singularity is practical only for small OAM numbers for two reasons: Automating the process of analyzing the fringe pattern, while in principle possible, is a non-trivial task. The clarity of fringes deteriorates in the presence of noise, especially for higher order OAM beams. We further note that the method is easily transferrable to the single-photon regime or for beams with non-integer OAM.}
\note{Knife-edge diffraction, as a method of identifying topological charge, has been demonstrated for electron vortices \cite{Guzzinati2014},}, and more generally, the method should be transferrable to any beams that can scatter, offering potential vortex measurements for media where other detection methods are costly or not yet available.




%
%

%

\begin{acknowledgments}
R.M.A. received funding via Fraunhofer CAP, award 322761-01.  SF-A acknowledges support through the QuantERA II Programme, with funding received via the EU H2020 research and innovation programme under Grant No. 101017733 and assoicated support from EPSRC under Grant No. EP/Z000513/1 (V-MAG). 
A.K.F. received funding via CAPES, Process number 88887.837237/2023-00. This study was financed in part by the Coordenação de Aperfeiçoamento de Pessoal de Nível Superior - Brasil (CAPES) - Finance Code 001.
\end{acknowledgments}

\subsection{Data Availability}*
The data that support the findings of this study are openly available from the Enlighten Repository at \note{to be included} 
reference number \note{to be included}.

\bibliography{main}

\end{document}